\documentclass[%
reprint,
superscriptaddress,
 amsmath,
 amssymb,
 aps,
pra
]{revtex4-1}
\usepackage{lipsum}

\usepackage{graphicx}
\usepackage{dcolumn}
\usepackage{bm}
\usepackage{epsfig}
\usepackage{epstopdf}

\usepackage{babel}
\usepackage[normalem]{ulem}
\makeatother
\usepackage{siunitx}
\usepackage{color}  
\begin{document}
\title{Characterization of soft X-ray echo-enabled harmonic generation free-electron
laser pulses in the presence of incoherent electron beam energy modulations}
\author{N. S. Mirian}
\email{najmeh.mirian@desy.de}

\affiliation{Deutsches Elektronen-Synchrotron DESY, Notkestraße 85, 22607 Hamburg, Germany}
\affiliation{Elettra-Sincrotrone Trieste S.C.p.A., 34149 Trieste, Italy}
\author{G. Perosa}
\email{giovanni.perosa@elettra.eu}

\affiliation{Elettra-Sincrotrone Trieste S.C.p.A., 34149 Trieste, Italy}
\affiliation{Università degli Studi di Trieste, Dipartimento di Fisica, Piazzale
Europa 1, Trieste, Italy}
\author{ E. Hemsing }
\affiliation{SLAC National Accelerator Laboratory, Menlo Park, California 94025,
USA}
\author{E. Allaria}
\affiliation{Deutsches Elektronen-Synchrotron DESY, Notkestraße 85, 22607 Hamburg, Germany}
\affiliation{Elettra-Sincrotrone Trieste S.C.p.A., 34149 Trieste, Italy}
\author{L. Badano}
\author{P. Cinquegrana}
\author{M. B. Danailov}
\author{G. De Ninno}
\affiliation{Elettra-Sincrotrone Trieste S.C.p.A., 34149 Trieste, Italy}
\author{L. Giannessi}
\affiliation{Elettra-Sincrotrone Trieste S.C.p.A., 34149 Trieste, Italy}
\affiliation{Istituto Nazionale di Fisica Nucleare - Laboratori Nazionali di Frascati
(INFN), 00044 Frascati, Rome, Italy}
\author{G. Penco}
\author{S. Spampinati}
\author{C. Spezzani}
\affiliation{Elettra-Sincrotrone Trieste S.C.p.A., 34149 Trieste, Italy}
\author{E. Roussel}
\affiliation{Universit\'e de Lille, CNRS, UMR 8523 - PhLAM - Physique des Lasers Atomes
et Molécules F-59000 Lille, France}
\author{P. R. Ribi\v{c}}
\author{M. Trovó}
\author{M.Veronese}
\affiliation{Elettra-Sincrotrone Trieste S.C.p.A., 34149 Trieste, Italy}
\author{S. Di Mitri}
\email{simone.dimitri@elettra.eu}

\affiliation{Elettra-Sincrotrone Trieste S.C.p.A., 34149 Trieste, Italy}
\begin{abstract}
Echo-enabled harmonic generation free-electron lasers (EEHG FELs)
are promising candidates to produce fully coherent soft x-ray pulses by virtue of efficient high harmonic frequency up-conversion from UV lasers.
The ultimate spectral limit of EEHG, however, remains unclear, because of the broadening and distortions induced in the output spectrum by residual broadband energy modulations in the electron beam. We present a mathematical description of the impact of incoherent
(broadband) energy modulations on the bunching spectrum produced by the microbunching instability through both the accelerator and the EEHG line. The model is in agreement
with a systematic experimental characterization of the FERMI EEHG
FEL in the photon energy range $130-210$ eV. We find that amplification of electron beam energy distortions primarily in the EEHG dispersive sections explains an observed reduction
of the FEL spectral brightness that is proportional to the EEHG
harmonic number.  Local maxima of the FEL spectral brightness and of the spectral
stability are found for a suitable balance of the dispersive sections'
strength and the first seed laser pulse energy. Such characterization
provides a benchmark for user experiments and future EEHG implementations designed to reach shorter wavelengths. 
\end{abstract}

\maketitle
\section{Introduction }

Free electron lasers (FELs) have enabled a new way for researchers to explore electronic dynamics at molecular and atomic scales via femtosecond
pulses, gigawatt peak powers, and tunable wavelengths in the range of
extreme ultraviolet to hard x-rays \cite{Milton2037}. Self-amplified
spontaneous emission (SASE) FELs generate a spiky spectrum and therefore
offer relatively poor longitudinal coherence \cite{Ackermann_2007,Emma_2010}.
Seeded FELs are, at present, the only devices producing stable pulses with good longitudinal
coherence at wavelengths now approaching the water window \cite{Allaria2012,Allaria2013}.

Echo-enabled harmonic generation (EEHG) was conceived as a seeding method with excellent high harmonic conversion efficiency to generate transform-limited radiation pulses down to soft x-rays \cite{Stupakov2009,Xiang2009,Hemsing2014,Hemsing2016,primoz2019,SXFEL}.
By utilizing two laser modulations and dispersive sections (DSs), a monochromatic
(coherent) energy modulation is imprinted on to the relativistic electron beam and transformed to a high harmonic density
modulation (see Fig.1). The beam then enters the undulator-radiator where the density-modulated (bunched) electrons
radiate coherently at wavelengths up to $\sim100$ times shorter than
that of the UV seeding lasers. With sufficient gain, the radiation can be amplified up to saturation.

Recently, the impact of phase variations in the seed lasers on EEHG performance
was investigated experimentally~\cite{mirian2020}, illustrating the capability to shape the EEHG FEL spectrum by tuning the second seed laser power and phase. The agreement between experimental data and the preceding theory~\cite{Stupakov:2011sd,Ratner2012,Hemsing2019} was obtained
in a condition in which energy non-uniformities of the electron beam
could be neglected. On the other hand, it is well know that energy distortions in the electron beam can impact the EEHG bunching spectrum (See, e.g., Refs.~\cite{Huang:2009zzd,PennPRSTAB2014,primoz2017}). Particularly relevant to this work, the impact of incoherent energy modulations on EEHG performance was discussed theoretically in \cite{PennNGLSTechnote35,Hemsing2017,Hemsing2018}.
Here, we examine the details of these studies by means of a extensive theoretical formulation of the evolution of energy and density modulations in EEHG. In particular, we examine and compare the measured FEL performance with an analytical model that includes incoherent modulations in the electron beam longitudinal phase space that develop from the early beam acceleration process through the final EEHG transformations.

\begin{figure*}
\includegraphics[width=0.9\textwidth]{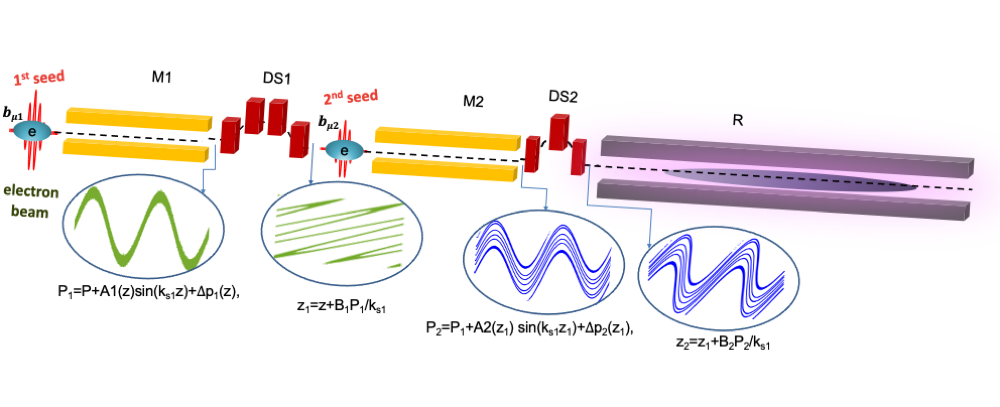}\caption{Main components of the EEHG scheme: first modulator (M1), strong first
dispersive section (DS1), second modulator (M2), weaker second dispersive
section (DS2). Equations  refer to quantities of the electron beam longitudinal phase space introduced in eq.(\ref{eq:coordinates}). In particular,  $\Delta p_{1,2}$ is the incoherent energy modulation (see eq.(\ref{eq:energyMBI})) and $b_{\mu1 ,\mu2}$ is the incoherent bunching factor (see eq.(\ref{eq:energyM})). After DS2, the nano-bunched electron beam travels into
the radiator (R) and emits coherent and powerful light pulse.}
\end{figure*}


It is known that in seeded FEL systems, uncontrolled energy structures
can lead to a broader FEL bandwidth and reduced peak spectral intensity. These energy
structures, accumulated during beam manipulation in the accelerator,
can hardly be removed completely. They can introduce extra frequencies
into the FEL gain bandwidth that deteriorate the longitudinal coherence promised by external seeding \cite{Hemsing2018,Perosa2020,marcus2019}.
Such structures are generally due to the build-up of beam collective effects such as coherent synchrotron
radiation (CSR) and longitudinal space charge (LSC)
during acceleration and compression, both contributing
to the so-called microbunching instability (MBI) \cite{Huang2002,SALDIN2002,Stupakov2002}
whose maximum gain, depending on the compression setting, is typically at final wavelengths $\lambda_{0}\gtrsim0.5\mu$m.
Here we report on results of a systematic investigation at the FERMI FEL operating in EEHG mode~\cite{primoz2019} where we find good agreement between theory and experimental data. These studies help to benchmark
the analytic model which thus provides a practical tool for the design and optimization of EEHG sources at even shorter wavelengths.

The paper is organized as follows. Section \ref{sec:theory} discusses the theory of MBI in EEHG. Section III presents the experiment results and
compares them with theory. Conclusions are reached in Section IV.

\section{Coherent and incoherent energy and density modulations}
\label{sec:theory}
\subsection{Theoretical Background}

The evolution of the electron beam longitudinal phase space through
the EEHG line in the presence of energy distortions is described by the following equations \cite{Hemsing2017,Hemsing2018},
: 
\begin{alignat}{1}
P_{1}= & P+A_{1}(z)\sin(k_{s1}z)+\Delta p_{1}(z),\nonumber \\
z_{1}= & z+B_{1}P_{1}/k_{s1},\label{eq:coordinates}\\
P_{2}= & P_{1}+A_{2}(z_1)\sin(k_{s2}z_{1})+\Delta p_{2}(z_{1}),\nonumber \\
z_{2}= & z_{1}+B_{2}P_{2}/k_{s1},\nonumber 
\end{alignat}
where $A_{1,2}(z)=\Delta E_{1,2}(z)/\sigma_{E}$ is the normalized
coherent energy modulation from seed lasers, and $B_{1,2}=k_{s1}R_{56}^{(1,2)}\sigma_{E}/E$
is the normalized energy dispersion in the chicanes, $E$ is the electron beam
mean energy and $\sigma_{E}$ is the RMS slice energy spread, and
$\Delta p_{1,2}(z)$ the energy distortions of the electron beam distribution.
The role of these terms is carefully investigated in the following.

In this description, $\Delta p_{1}$ represents any
energy structure accumulated in the electron beam up to the entrance of the first EEHG chicane $B_1$. $\Delta p_{2}$ is used to capture the integrated
effect of CSR from $B_1$ and of LSC in second modulator.
They can be expressed as the superposition of monochromatic modulations
of different amplitudes \cite{Ratner2015,Hemsing2018}: 
\begin{eqnarray}
\Delta p_{1,2}(z) & = & \sum_{\mu=0}^{\infty}p_{1,2}(k_{\mu})\sin(k_{\mu}z+\phi_{1,2\mu}),\label{eq:energyMBI}
\end{eqnarray}
where $\phi_{1,2\mu}$ is a random phase.

When the energy distortions $\Delta p_{1,2}$ are
ignored and in the assumption of uniform lasers $A_{1,2}(z)=A_{1,2}$ (i.e., seed durations much longer than bunch duration),
the Fourier transform of the electron beam density distribution
- the so-called bunching factor - can be calculated at the exit of
the second EEHG modulator according to \cite{Stupakov2009}: 
\begin{equation}
\bar{b}_{n,m}(k_{E})=e^{-\zeta_{E}^{2}/2}J_{n}(-\zeta_{E}A_{1})J_{m}(-a_{E}A_{2}B_{2}),
\label{eq:EEHGBunchingF}
\end{equation}
where $a_{E}=n+mk_{s2}/k_{s1}$ is the harmonic number with integer
numbers $n$ and $m$. The EEHG wave number is $k_{E}=a_{E}k_{s1}$,
and $\zeta_{E}=nB_{1}+a_{E}B_{2}$. This factor is known to help characterize
the EEHG performance and is optimized approximately at $\zeta_{E}=j'_{n,1}/A_1$ where $j'_{n,1}$ is the first root of $J'_n$.

If the energy distortion $\Delta p_{2}$ and second laser $A_2$ are sufficiently slowly-varying longitudinally that we can approximate their functional dependence as $z_{1}\approx z$, then the EEHG bunching factor close to the harmonic peak becomes \cite{Hemsing2018,Hemsing2019}:
\begin{widetext}
\begin{eqnarray}
b_{n,m}(k) & = & e^{-\frac{1}{2}\big(\zeta_{E}+\frac{k-k_{E}}{k_{1s}}B\big)^{2}}\int_{-\infty}^{+\infty}dz\,f(z)J_{m}\bigg[-\frac{k}{k_{s1}}B_{2}A_{2}(z)\bigg]\times\nonumber \\
 &  & J_{n}\bigg[-\bigg(\zeta_{E}+\frac{k-k_{E}}{k_{s1}}B\bigg)A_{1}(z)\bigg]\times\label{eq:complestBF}  \nonumber \\
 &  & e^{i(-\zeta_{E}\Delta p_{1}(z)-a_{E}B_{2}\Delta p_{2}(z)+(k-k_{E})z)},\label{eq:bunching_complete}
\end{eqnarray}
\end{widetext}
where $f(z)$ is electron beam density distribution function, and $B=B_1+B_2$. In the first two gain lengths in the radiator (R in Fig.1) the intensity of the FEL radiation is estimated to grow like $\propto z^2|b_{n,m}(k)|^2$. In the limit of negligible slippage the final radiation spectral pulse properties are given by $|b_{n,m}(k)|^2$. 
Thus, by virtue of the general expression for the energy modulations given in (\ref{eq:energyMBI}), this equation can be used to quantify the spectral effect of broadband energy modulations induced by MBI on the FEL output. 
\subsection{Bunching Phase}
The $z$-dependent additional bunching phase due to electron beam energy distortions in Eq.(\ref{eq:bunching_complete}) is: 
\begin{equation}
\psi(z)=-\zeta_{E}\Delta p_{1}(z)-a_{E}B_{2}\Delta p_{2}(z).
\label{eq:additional_phase}
\end{equation}
From this, one can obtain the moments of the spectral bunching distribution and gain insight into the relative magnitude of the contributions from $\Delta p_{1,2}$~\cite{Hemsing2018}.

MBI-induced energy modulations accumulated up to the exit of the first
modulator, $\Delta p_{1}$, are multiplied by the small scaling parameter
$|\zeta_{E}|\lesssim1$. Linear,
quadratic, and sinusoidally-shaped initial modulations were investigated
in \cite{Hemsing2017}, where it was shown that the smallness
of $\zeta_{E}$ accounts for the insensitivity of the EEHG bunching
spectrum to small initial perturbations. However, as discussed in \cite{Hemsing2018}, energy modulations
$\Delta p_{2}$ that develop between the EEHG chicanes are multiplied
by the much larger factor $a_{E}B_{2}\approx m/A_{2}\gg1$, and therefore can have a noticeable impact on the final bunching spectrum at
high harmonics.

The RMS bandwidth
of $\left|b_{n,m}(k)\right|^{2}$ in Eq.(\ref{eq:complestBF}) is $\sigma_{k}^{2}=\sigma_{ks}^{2}+\sigma_{\psi^{\prime}}^{2},$
where $\sigma_{ks}$
is the transform-limited (TL) bandwidth and $\sigma_{\psi^{\prime}}^2=\left\langle \left[\psi^{\prime}-\left\langle \psi^{\prime}\right\rangle \right]^{2}\right\rangle $
is the bandwidth due to the nonlinear phase structure, where brackets denote integration over the $z$-dependent amplitudes in the integrand in (\ref{eq:bunching_complete}). 
Assuming the bunching longitudinal envelope is determined by the second seed laser and that it is a TL Gaussian pulse, the relative bandwidth in the case of optimized bunching absent MBI can be approximated as \cite{Hemsing2019}:
\begin{equation}
\bar{\sigma}_{ks}^{2}=\frac{4\bar{\sigma}_{ks2}^{2}}{3m^{4/3}},
\label{eq:seedeffect}
\end{equation}
where $\bar{\sigma}_{ks2}$ is the relative second seed bandwidth. 

Inserting Eq.~(\ref{eq:energyMBI}) for broadband energy distortions into the phase in (\ref{eq:additional_phase}), the instantaneous spatial bunching
frequency is $k_{z}=k_{E}+\psi^{\prime}(z)$, where: 
\begin{widetext}
\begin{equation}
\psi^{\prime}(z)=-\zeta_{E}\sum_{\mu=0}^{\infty}p_{1}(k_{\mu})\,k_{\mu}\cos(k_{\mu}z+\phi_{1\mu})-a_{E}B_{2}\sum_{\mu=0}^{\infty}p_{2}(k_{\mu})\,k_{\mu}\cos(k_{\mu}z+\phi_{2\mu})\label{eq:frequency_fluctuation}
\end{equation}
is the $z$- derivative of the additional phase. The mean bunching frequency is then $\langle k_z\rangle$. Thus, $\langle \psi^{\prime}\rangle$ gives the spectral shift from $k_E$, and $\sigma_{\psi^{\prime}}$ gives the excess bandwidth due to the distortions. 
Assuming that the characteristic MBI wavelengths are small compared the length of the bunching envelope (e.g, $k_{\mu} \gg \sigma_{ks}$) and that the individual phases $\phi_{1,2\mu}$ are uncorrelated over $\mu$, bandwidth of $|b_{n,m}(k)|^{2}$ is therefore:
 
\begin{equation}
\sigma_{k}^{2}=\sigma_{ks}^{2}+\sum_{\mu=0}^{\infty}\left[\frac{\zeta_{E}^{2}}{2}(p_{1}(k_{\mu})\,k_{\mu})^{2}+\frac{\left(a_{E}B_{2}\right)^{2}}{2}(p_{2}(k_{\mu})\,k_{\mu})^{2}\right].\label{eq:FEL_spectrum}
\end{equation}
\end{widetext}
\subsection{Bunching Amplitude}
Similarly to the phase, the bunching factor in Eq.(\ref{eq:bunching_complete}) for generic energy distortions is here specialized for MBI-induced energy modulations described in Eq.(\ref{eq:energyMBI}). It becomes:\\
\begin{widetext}
\begin{eqnarray}
b_{n,m}(k) & = & \int \chi(z, k)\,e^{-i(z(k-k_{E})+\psi(z))}dz\nonumber \\
 & = & \int dz\,\chi(z,k)\,\prod_{\mu=0}^{\infty}\sum_{l_{1}=-\infty}^{\infty}\sum_{l_{2}==-\infty}^{\infty}J_{l_{1}}\left(-\zeta_{E}p_{1}(k_{\mu})\right)\,J_{l_{2}}\left(-a_{E}B_{2}p_{2}(k_{\mu})\right)\times\label{eq:bunching_com}\\
 &  & e^{-iz\left[k_{\mu}(l_{1}+l_{2})-(k-a_{E}k_{1s})\right]}e^{-i(l_{1}\phi_{1\mu}+l_{2}\phi_{2\mu})},\nonumber 
\end{eqnarray}
where 
\begin{equation}
\chi(z, k)=e^{-\frac{1}{2}\big(\zeta_{E}+\frac{k-k_{E}}{k_{1s}}B\big)^{2}}
J_{m}\bigg[-\frac{k}{k_{s1}}B_{2}A_{2}(z)\bigg] J_{n}\bigg[-\bigg(\zeta_{E}+\frac{k-k_{E}}{k_{s1}}B\bigg)A_{1}(z)\bigg]\, f(z).\label{eq:chi}
\end{equation}
With the definition of the bunching spectrum $b_{n,m}(k)$, we can
now quantify the presence of sidebands and/or of a broader spectral
pedestal in EEHG.
The EEHG bunching amplitude evaluated for $k=a_{E}k_{1s}$
can be calculated when $l_1=-l_2$, so that: 
\begin{eqnarray}
b_{n,m}(a_{E}k_{1s})& = &\grave{b}_{n,m} \prod_{\mu=0}^{\infty}\sum_{l_{1}=-\infty} ^ {\infty} (-1)^{l_1}J_{l_{1}}\left(-\zeta_{E}p_{1}(k_{\mu})\right)\,J_{l_{1}}\left(-a_{E}B_{2}p_{2}(k_{\mu})\right) e^{-il_{1}(\phi_{1\mu}-\phi_{2\mu})}, \label{eq:mainpeak2}
\end{eqnarray}
where $\grave{b} _{n,m}$ is the z-integration of Eq.(\ref{eq:chi}) for $k=a_Ek_{1s}$ and demonstrates the bunching factor when MBI is absent. In above equation we use the Bessel function relationship for integer $\nu$ value $J_{-\nu}(x)=(-1)^\nu J_{\nu}(x)$.
In the case of long seed lasers (i.e $A_{1,2}(z)=A_{1,2}$) and a uniform electron beam, it is easy to see that $\grave{b} _{n,m}=\bar{b}_{n,m}$.
Note that the bunching is suppressed at the roots of the two
Bessel functions. Assuming that the arguments of $J_{l_1}$ in Eq.(\ref{eq:mainpeak2}) are less than 1, the high order of Bessel functions can be ignored and the leading term can be expanded around $0$. In doing so, the bunching factor can
be simplified to: %
\begin{eqnarray}
b_{n,m}(a_{E}k_{1s}) & \thickapprox &\grave{b}_{n,m} \prod_{\mu=0}^{\infty} J_0\left(-\zeta_{E}p_{1}(k_{\mu})\right)\,J_0\left(-a_{E}B_{2}p_{2}(k_{\mu})\right) \\ \nonumber
 & \thickapprox & \grave{b}_{n,m}\prod_{\mu=0}^{\infty}\left[1-\frac{1}{4}\left(\left(\zeta_{E}p_{1}(k_{\mu})\right)^{2}+\left(a_{E}B_{2}p_{2}(k_{\mu})\right)^{2}\right)\right].\label{eq:mainpeak}
\end{eqnarray}
\end{widetext}

\subsection{Modeling the Microbunching Instability}

For numerical modeling, the expressions of the perturbed bunching factor can now be made explicit
for the MBI-induced broadband energy modulations accumulated up to
the second EEHG DS. At typical frequencies $k_{\mu}\gg1/\sigma_{z}$,
we can write \cite{Hemsing2018,Huang2010}: 
\begin{equation}
p_{1,2}(k_{\mu})=4\pi b_{\mu 1,\mu 2}(k_{\mu})\frac{I}{I_{A}}\frac{Z_{1,2}(k_{\mu})}{Z_{0}\sigma_{\gamma}}L_{1,2},\label{eq:energyM}
\end{equation}
where $p_{1,2}$, $b_{\mu1,\mu2}(k_{\mu})$, and $Z_{1,2}(k)$ are the broadband
energy modulation, broadband bunching factor, and LSC impedance per unit length in the first and second modulator of length $L_{1,2}$. $I$ and $I_{A}=17045A$
are the bunch peak current and the Alfven current, and $Z_{0}=377\Omega$
is the free space impedance. The in-vacuum LSC impedance $Z(k_{\mu})$
through the modulator is \cite{Huang2004,Huang2010}: 
\begin{equation}
Z(k_{\mu}/C)=\frac{iZ_{0}k_{\mu}}{4\pi\gamma_{z}^{2}C}\left[1+2\ln\bigg(\frac{\gamma_{z}}{k_{\mu}r_{b}}\bigg)\right],\label{eq:impedance}
\end{equation}
with the effective beam transverse size $r_{b}=0.8735\sqrt{\epsilon_{x}\beta_{x}+\epsilon_{y}\beta_{y}}$,
where $\epsilon_{x,y}$ and $\beta_{x,y}$ are emittance betatron functions in x and y directions,
$\gamma_{z}=\gamma/\sqrt{1+K_{u}^{2}/2}$ is the longitudinal Lorentz
factor inside an undulator with the (peak) untapered undulator parameter $K_u$, and $C$ is compression factor.

The MBI-induced energy modulation $\Delta p_{1,2}$ in Eq.(\ref{eq:energyMBI})
are calculated numerically by means of a comprehensive linear gain model of the
instability from beam injection into the accelerator up to the undulator
line, including longitudinal energy-dispersion function ($R_{56}$)
and CSR in the magnetic compressor, LSC and intrabeam scattering,
and beam heating at low energy \cite{DiMitri2020,DiMitri2021}. The
model starts from a shot-noise like initial bunching factor and provides
the bunching factor at any point along the line as well.

In the simplified case of single-stage beam compression and assuming linear gain regime of the instability, the amplification of density modulation, or gain
\cite{Huang2002,Stupakov2002}, in the presence of an arbitrary incoming
energy distribution $V(P_{0})$ \cite{Huang2004}, goes like: 
\begin{eqnarray}
G & = & \left|\frac{b_{\mu f}}{b_{\mu0}}\right|\backsimeq\frac{I}{\gamma I_{A}}\left|k_{f}R_{56}\intop_{0}^{L}ds\frac{4\pi Z(k_{\mu0},s)}{Z_{0}}\right|\nonumber \\
 &  & \times\int dP_{0}V(P_{0})e^{-ik_{\mu f}R_{56}P_{0}},\label{eq:MBIgain}
\end{eqnarray}
where $k_{\mu f}=k_{\mu0}/(1+hR_{56})$, $k_{\mu0}$ and $k_{\mu f}$
are modulation wavelength before and after the beam compression, and $h$ is the initial beam energy chirp.

Equation (\ref{eq:MBIgain}) suggests that the gain can be reduced
by energy Landau damping, i.e., by increasing the beam uncorrelated
energy spread at relatively low beam energies. Indeed, this is now
accomplished at several FEL facilities through the laser heater (LH)
system \cite{SALDIN2004,PAL-XFEL,FERMI_LH,Huang2010}, whose accurate
control has become a tool to optimize the FEL spectral brightness in
the presence of MBI. 
\section{Echo-Enabled Harmonic Generation Measurements}

\subsection{Bandwidth Enlargement and Central Frequency Fluctuation}

The EEHG experiment was conducted with an electron beam accelerated
through the FERMI linac to the final energy of $E$=1.32 GeV. The beam
normalized emittance measured in front of the undulator amounts to
approximately 1 mm mrad in both transverse planes. The electron bunch
is compressed by a factor $C\sim$10 to reach a final peak current
in the core of $I$=700 A.

\begin{figure*}[!ht]
\includegraphics[width=12cm]{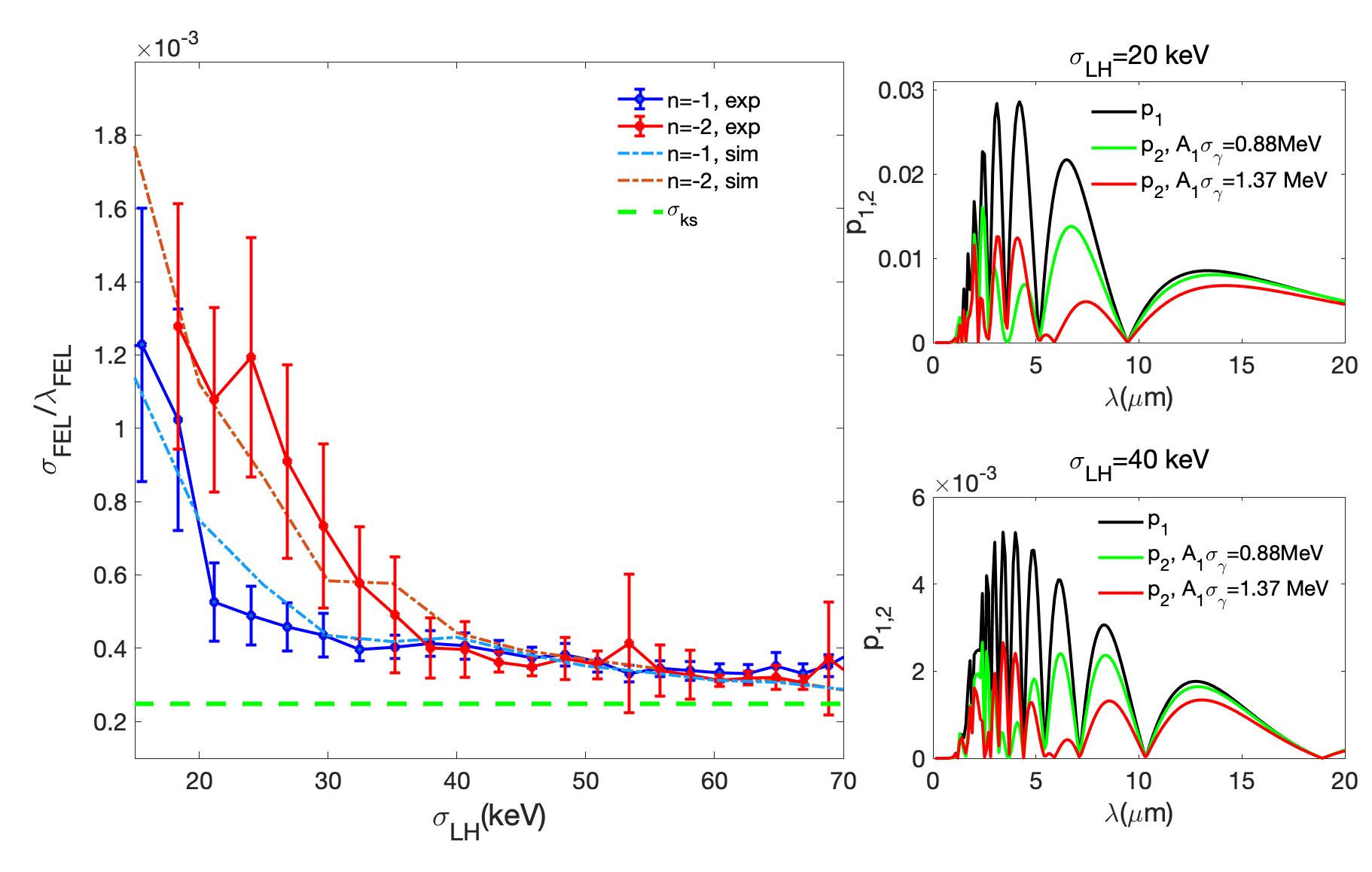}\caption{Left: Relative FEL RMS spectral bandwidth vs. LH-induced energy spread.
Blue and red lines are experimental data (solid) and theoretical prediction
(dashed-dotted, Eq.(\ref{eq:FEL_spectrum})) for $n=-1$ and $n=-2$, respectively.
EEHG is tuned at $\lambda_{FEL}=8.8$ nm. The dashed green line is
from Eq.(\ref{eq:seedeffect}). In case of $n=-1$, the first seed
energy is $8.3\,\mu J$ ($A_{1}\sigma_{E}=0.88MeV$) and for
$n=-2$, it is $20\,\mu J$ ($A_{1}\sigma_{E}=1.37MeV$). Right:
calculated energy modulation amplitude $p_{1}(\lambda_{\mu})$ and
$p_{2}(\lambda_{\mu})$ from MBI modeling for $\sigma_{LH}=$20 kev
(top) and 40 kev (bottom) . \label{fig:FEL-spectrum}}
\end{figure*}

Figure \ref{fig:FEL-spectrum}-left plot shows a scan of the RMS spectral
bandwidth of the FEL at harmonic $a_{E}=30$ of a UV seed laser ($\lambda_{s}=264.54$
nm), as a function of the LH-induced energy spread. The dispersion
of the first EEHG dispersive section was set to $R_{56}^{(1)}=2.25$
mm. The blue curve shows the EEHG emission in $n=-1$ configuration or
$R_{56}^{(2)}=75\ \mu$m; the red curve is for $n=-2$ or $R_{56}^{(2)}=145\ \mu$m
($n$ defined in Eq.(\ref{eq:EEHGBunchingF})). The error bars reflect
the RMS fluctuation of experimental data collected over a series of
20 consecutive shots at 10 Hz machine repetition rate. The experimental data are
compared with the theoretical bandwidth predicted by Eq.(\ref{eq:FEL_spectrum})
for $n=-1$ and $n=-2$, illustrated by the the dashed-dotted blue and red
line, respectively. For comparison, the green dashed line represents
the bandwidth for optimized bunching absent MBI, Eq.(\ref{eq:seedeffect}),
assuming TL seed laser pulses with a FWHM bandwidths of 2.01 nm and 1.35 nm, respectively.

The spectral content of the energy
distortion amplitudes predicted by the MBI model for two different LH energy spread settings is shown in the right Fig.~\ref{fig:FEL-spectrum} subplots. The integrated impact of these distortions matches well the  measured FEL bandwidth, which is substantially reduced for a LH-induced energy spread $\geq30$ keV. The model allows us to explain
the observations on the basis of MBI-induced energy modulations augmented
by the first EEHG dispersive section, where $p_{2}(\lambda_{\mu})$ in absent of the first seed results always larger than $p_{1}(\lambda_{\mu})$. 

The different MBI sensitivity of the EEHG bandwidth for the cases
$n=-1$ and $n=-2$ is explained by means of Eq.(\ref{eq:FEL_spectrum}).
On the one side, $p_{1}(\lambda_{\mu})$ is multiplied by the EEHG
scaling factor, which therefore can be modified to change the sensitivity
of the final bunching to the electron beam energy perturbations coming
from the accelerator. On the other side, $p_{2}(\lambda_{\mu})$ is
multiplied by $a_{E}B_{2}$, with $|B_2|\approx nB_1/a_{E}$, such that a higher
value of $|n|$ forces larger values $R_{56}^{(2)}$ of the second dispersion section.

\begin{figure}
\includegraphics[width=7.5cm,height=6cm]{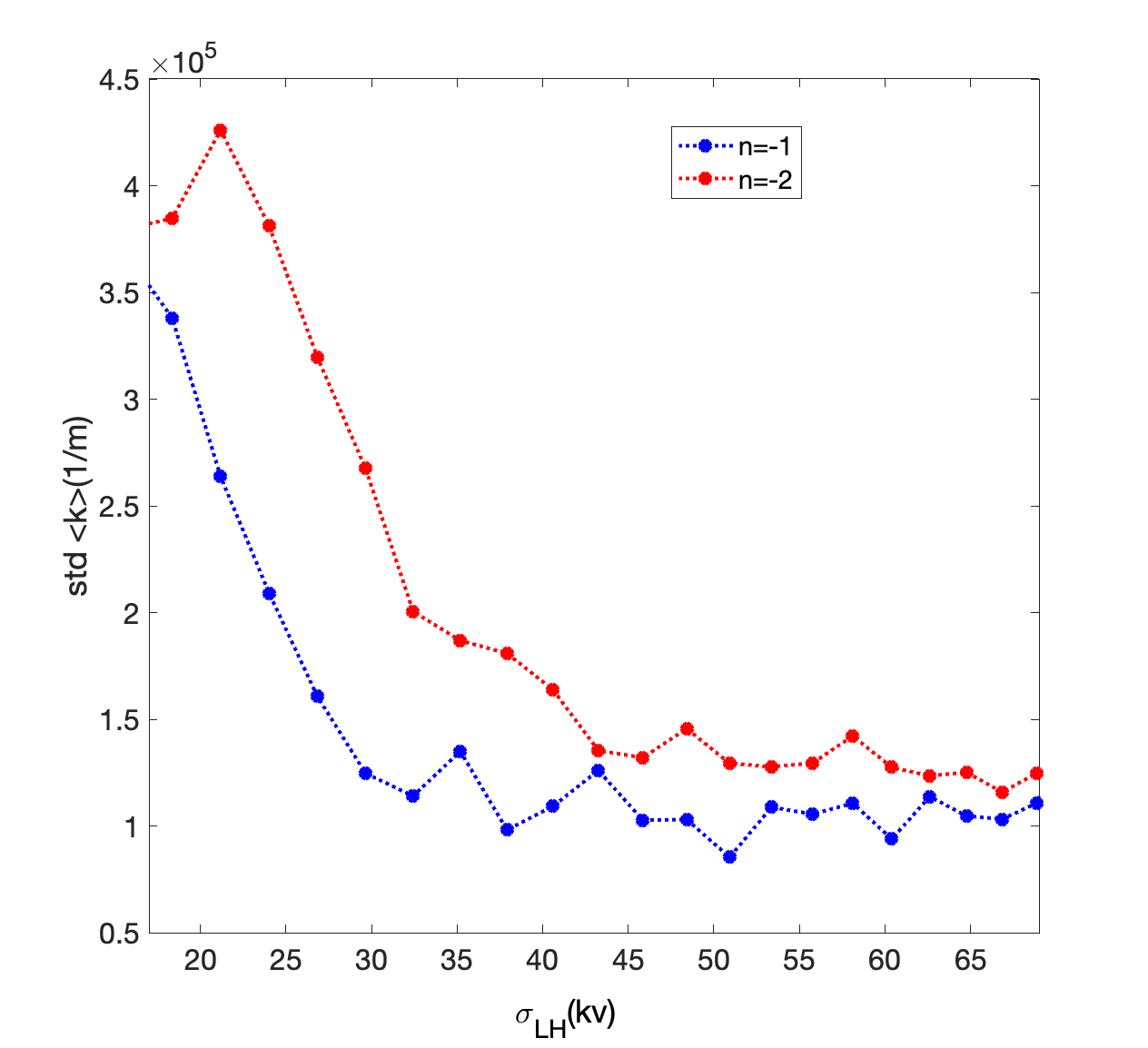}\caption{Comparison of standard deviation of frequency fluctuations of 50 shots
$n=-1$ (blue) and 100 shots $n=-2$ (red) configurations in EEHG experiment
respect to the different induced LH energy spread. EEHG harmonic
is 30 ( $\lambda_{FEL}=8.8$ nm). The FEL parameters are same as in
Fig.\ref{fig:FEL-spectrum}.}
\label{fig:frequency_STD} 
\end{figure}

Equation (\ref{eq:frequency_fluctuation}) suggests that, by virtue
of larger values of $B_{2}$ in the presence of MBI, frequency fluctuations
in the configuration $n=-2$ for fixed $\zeta_{E}$, should be larger
than in $n=-1$. Figure \ref{fig:frequency_STD} compares the range of the frequency fluctuation by showing the standard
deviation (std) of 50 single shots of $n=-1$ (blue) and 100 shots
of $n=-2$ (red) configurations in EEHG experiment at $\lambda_{FEL}=8.8 $ nm
respect to the different induced LH energy spread. In other words, this figure shows the range of $\langle k_z\rangle$ fluctuations for different level of incoherent energy modulation. 
The larger fluctuations seen with the $n=-2$ setting align with expectations.

\begin{figure}
\includegraphics[width=8cm]{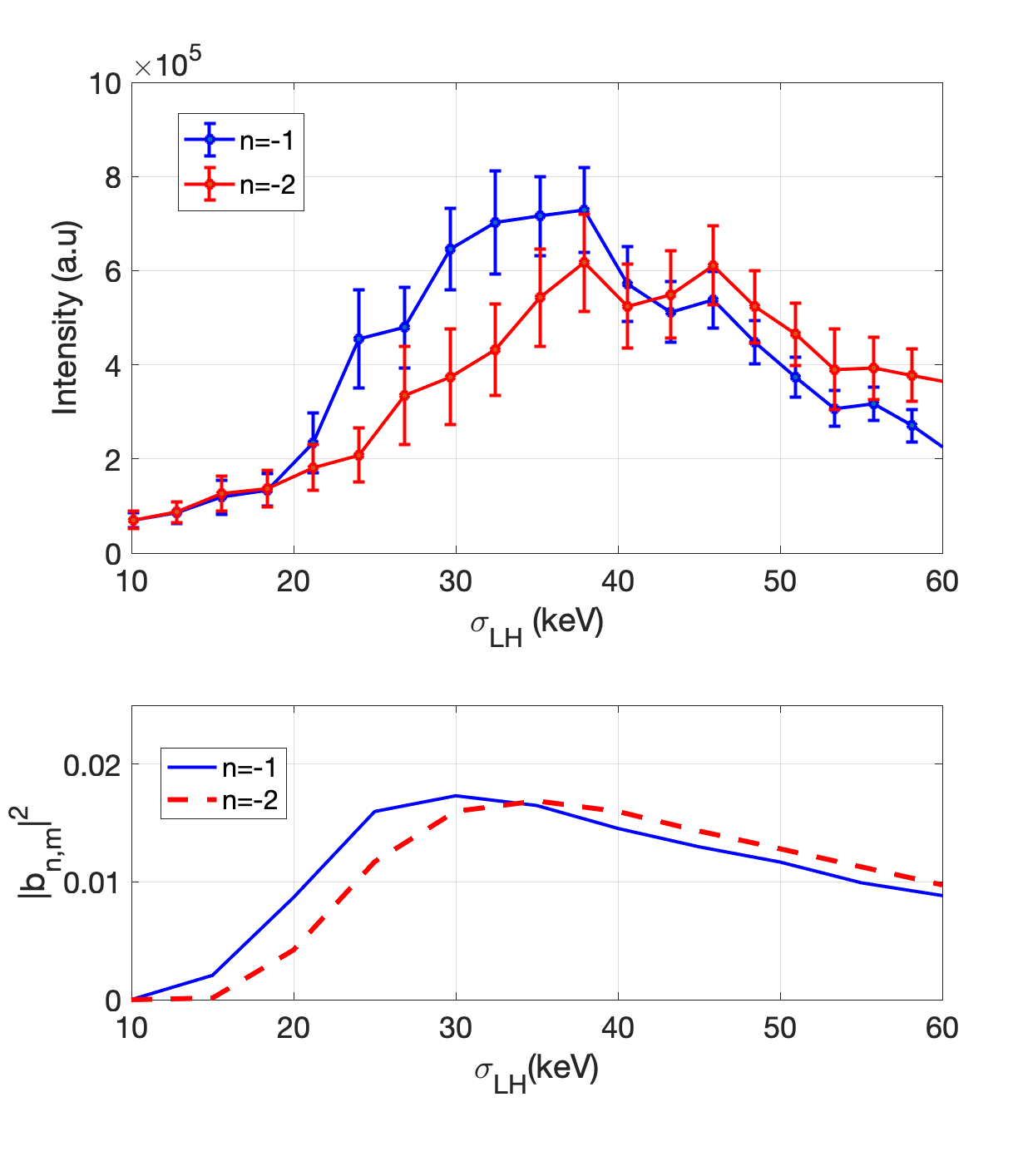} \caption{Top: FEL intensity vs. LH-induced energy spread. The FEL parameters are same as in Fig.\ref{fig:FEL-spectrum}. Bottom:
square bunching factor. The calculated value of $p_{1}(\lambda_{\mu})$
and $p_{2}(\lambda_{\mu})$ in Fig.\ref{fig:FEL-spectrum} are used
to evaluate bunching factor.
\label{fig:FELintensity_vsLH}}
\end{figure}

\subsection{Pulse Intensity Reduction}

Figure \ref{fig:FELintensity_vsLH}-top plot shows the the maximum measured
FEL intensity for $n=-1$ (blue line) and $n=-2$ (red line). As mentioned the FEL intensity scales with $|b_{n,m}(k_E)|^2$. The bottom plot shows the values calculated from Eq.(\ref{eq:mainpeak}). The equation shows that when
the MBI gain is suppressed by large LH pulse energies, the product
function $\Gamma=\prod_{\mu=0}^{\infty}\,J_{0}\left(-\zeta_{E}p_{1}(k_{\mu})\right)\,J_{0}\left(-a_{E}B_{2}p_{2}(k_{\mu})\right)\approx\prod_{\mu=0}^{\infty}\left[1-\frac{1}{4}\left(\left(\zeta_{E}p_{1}(k_{\mu})\right)^{2}+\left(a_{E}B_{2}p_{2}(k_{\mu})\right)^{2}\right)\right]$
tends to 1. Likewise, when the MBI is more pronounced at low LH pulse
energies, the product function approaches zero. At the same time,
owing to the large LH-induced energy spread (larger than 40 keV),
the FEL gain is diminished and therefore the FEL intensity is reduced.
We note that in the $\Gamma$ function, $p_{2}(\lambda_{\mu})$ is
multiplied by $a_{E}B_{2}$, which explains the different behaviour
of the function for $n=-1$ and $n=-2$, in agreement with the experimental
observation.

\subsection{Impact of first Seed Laser }

It is well-known that in the processes of harmonic emission driven
by an external laser, the seed laser-induced energy modulation has
to exceed the uncorrelated energy spread
of the beam at the undulator entrance. Moreover, the EEHG bunching becomes less sensitive to MBI with increased laser modulations. This leads to the question
if and to which extent the seeding laser pulse energy could be increased
in order to counteract the effect of MBI, before preventing any further
lasing by exceeding the FEL normalized energy bandwidth.

\begin{figure*}
\includegraphics[width=0.9\textwidth]{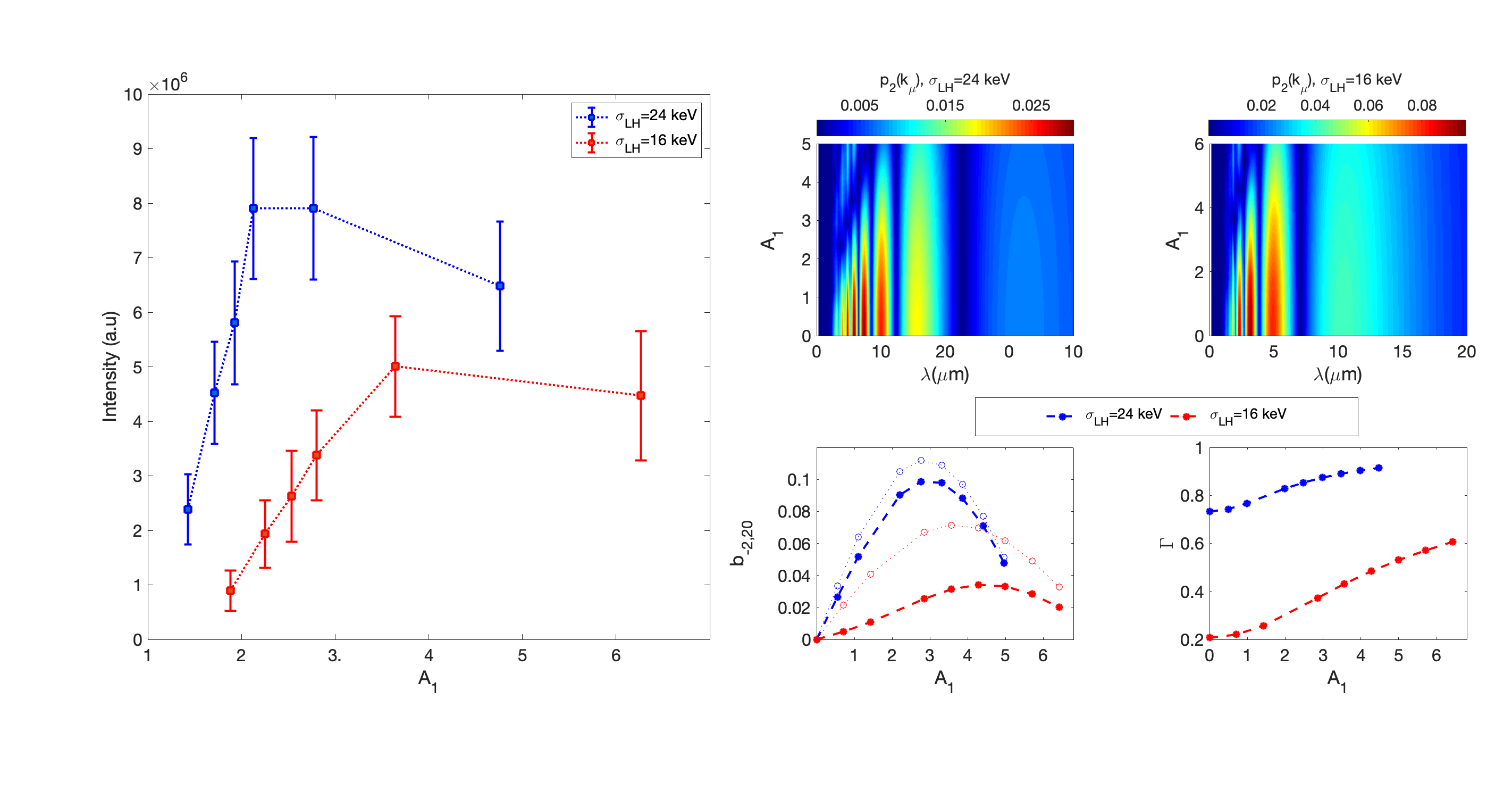} \caption{Left: FEL intensity vs. first seed laser pulse energy. Blue and red
lines are experimental data for LH-induced energy spread $\sigma_{LH}=24,\ keV$
and $\sigma_{LH}=16,\ keV$. Right-top row: beam energy modulation
amplitude as function of the (compressed) MBI modulation wavelength
and first seed coherent energy modulation amplitude (seed energy),
at the exit of the second modulator. Right-bottom row: bunching factor
and product function versus first seed coherent energy modulation
amplitude. Dotted lines with circles show $\grave{b}_{-2,20}$;
dashed lines with stars show $b_{-2,20}$ in Eq.(\ref{eq:mainpeak2}). The two
sub-plots refer to the LH inducing 24 keV and 16 keV RMS energy spread,
respectively. \label{fig:FEL-intensityvsseed}}
\end{figure*}

To answer this question, Fig.\ref{fig:FEL-intensityvsseed}-left plot
illustrates the FEL intensity recorded as function of the first seed
laser pulse energy, for two values of the LH pulse energy. Since the
bunching is more sensitive to the coherent energy modulation in the
first modulator at higher harmonics, the experiment was done for harmonic
18. The beam energy was 1.1 GeV and the compression factor about 7,
for approximately 550 A in the bunch core. EEHG was set in configuration
$n=-1$ ($R_{56}^{(1)}=1.9$ mm and $R_{56}^{(2)}=98\:\mu$m).

The figure shows that, once the FEL emission is built up for the seed
laser pulse energy of $\sim15$-$20\ \mu$J, the intensity is weakly
affected by even larger seed energies. In particular, a stronger seed
laser is not able to recover the intensity reduction due to a weaker
heating effect. The right plots provide the theoretical explanation
of the experimental data. They show the spectrum of broadband energy
modulation at the exit of the second modulator for different coherent
energy modulations from the first seed laser, at two LH pulse energies.
The contour plots are generated by inserting the energy distribution
modified by the first seed laser into Eq.(\ref{eq:MBIgain}), which
allows us to calculate the MBI gain at the exit of the first DS. Using
such spectral gain function as an input to Eq.(\ref{eq:energyM}), the
spectrum of the MBI-induced broadband energy modulation at the exit
of the second modulator is eventually derived as function of the first
seed pulse energy. An increase of the first coherent modulation is
effective in removing incoherent energy modulations at initial wavelengths
shorter than 10 $\mu$m, or $\sim1\ \mu$m at the entrance of the
undulator. However, the effect becomes negligible at immediately longer
wavelengths.

We plugged these two sets of energy modulation and different first
seed energies into Eq.(\ref{eq:EEHGBunchingF}) and Eq.(\ref{eq:mainpeak}),
thus calculated the bunching factor ($b_{-2,20}$) and the product
function ($\Gamma$), see bottom plots. It is shown that, while a
strong beam heating is able to shift the product function to 1 or so,
an increase of $A_{1}$ is not able to recover a unitary product function
(compare red starts and blue stars for $\sigma_{LH}=16$ keV and $\sigma_{LH}=24$
keV, respectively). The second plot of the second row compares the
bunching factor with (dashed lines with 
stars, Eq.(\ref{eq:mainpeak})) and without MBI (doted lines with
circle, Eq.(\ref{eq:EEHGBunchingF})). The plot illustrates the degradation
of the bunching factor by MBI, at different coherent energy modulation
amplitudes from the first seed. Finally, we find that the measured
FEL intensity (left plot) at the two LH pulse energies is in agreement
with the theoretical behaviour of the bunching factor: by increasing
the first seed energy, the FEL intensity grows up, to eventually fall
down for exceeding seeding energies.\\
%
%
\section{Conclusions}

Electron beam imperfections play a significant role in determining
the spectral bandwidth and the pulse intensity of EEHG FEL emission.
While EEHG is predicted to be more robust than other external seeding schemes to energy distortions that occur upstream, it is also anticipated that distortions that occur between the EEHG chicanes can significantly impact the FEL spectrum. This due in part to the combination of a strong magnetic chicane in the EEHG set up (compared to HGHG, for example)
and longitudinal space charge forces acting
through the second modulator that together can amplify incoherent microbunching generated upstream in the accelerator.
Here, the role played by the instability in different EEHG configurations was illustrated with experimental data, and good agreement was found with start-to-end semi-analytical results
for the perturbed bunching factor. As such, the model turns out to be a practical tool for 
the design and optimization of short wavelength EEHG sources.
Notably, the higher the EEHG harmonic jump,
the more sensitive the FEL sprectral brightness is to incoherent
energy modulations through the second modulator. Moreover, different
balances of the strength of the two dispersive sections change the impact of MBI onto
the FEL spectrum and intensity, with smaller value of the $|n|$-factor
less sensitive to incoherent energy modulations, as predicted.
Finally, attempts to maximize the FEL intensity with a stronger coherent energy
modulation from the first seed laser pulse were successful only for
moderate or strong beam heating in the first stages of acceleration.
This identifies a limitation in recovering optimal EEHG performance
through the seed laser pulse energy. Further, it suggests a careful control of the instability in the accelerator,
and a consequent optimization of the EEHG set up in the presence of relatively large heating levels.
\appendix
\section{}

The derivation of equations (\ref{eq:mainpeak}) 
follows the strategy proposed in \cite{Hemsing2017}. Starting from
the general expression (\ref{eq:complestBF}) in presence of energy
distortions, we retain only the lowest order contribution near the
harmonic: 
\begin{eqnarray}
b(k) & \approx & \int_{-\infty}^{+\infty}dz\chi(z,k)e^{i(k-k_{E}+n\psi_{1}+m\psi_{2})z}\times\nonumber \\
 &  & e^{i(-\zeta_{E}\Delta p_{1}(z)-a_{E}B_{2}\Delta p_{2}(z_{1}))z}.\label{passage1}
\end{eqnarray}

From here on, we will assume that $\Delta p_{1}$ is small enough
not to alter significantly the phase space distribution after the
first dispersive region. Also, we will further simplify our calculation
taking $\Delta p_{2}(z_{1})=\Delta p_{2}(z)$. This assumption is
true as long as $\Delta p_{2}(z_{1})$ is a sufficiently slowly-varying
function.

Using the definition of $\Delta p_{1,2}$ given in eq.(\ref{eq:coordinates}),
and the identity 
\begin{equation}
e^{ix\sin{\theta}}=\sum_{n=-\infty}^{\infty}J_{n}(x)e^{in\theta}.
\end{equation}
it is possible to expand the phase associated to broadband energy
modulations, 
\begin{widetext}
\begin{eqnarray*}
e^{\left[-i\zeta_{E}\Delta p_{1}(z)\,z\right]} & = & \prod_{\mu=0}^{\infty}\sum_{l_{1}=-\infty}^{\infty}\exp\left[il_{1}k_{\mu}z+il_{1}\phi_{1\mu}\right]J_{l_{1}}\left[-\zeta_{E}p_{1}\left(k_{\mu}\right)\right],\\
e^{\left[-ia_{E}B_{2}\Delta p_{2}(z)\,z\right]} & = & \prod_{\mu=0}^{\infty}\sum_{l_{2}=-\infty}^{\infty}\exp\left[il_{2}k_{\mu}z+il_{2}\phi_{2\mu}\right]J_{l_{2}}\left[-a_{E}B_{2}p_{2}\left(k_{\mu}\right)\right],
\end{eqnarray*}
\end{widetext}

and equation (\ref{passage1}) reduces to eq.(\ref{eq:bunching_com}).\\

The maximum EEHG bunching amplitude evaluated for $k=a_{E}k_{1s}$
is: 
\begin{equation}
b_{n,m}(a_{E}k_{1s})=\grave{b}_{n,m}\prod_{\mu=0}^{\infty}\,J_{0}\left(-\zeta_{E}p_{1}(k_{\mu})\right)\,J_{0}\left(-a_{E}B_{2}p_{2}(k_{\mu})\right).
\end{equation}
Assuming that the energy distortions amplitudes, namely $p_{1,2}$,
are small compared to the energy modulations induced by the seed laser,
we can expand the Bessel functions around 0. For a generic $\alpha$,
we have 
\begin{equation}
J_{\alpha}(x)=\sum_{m=0}^{\infty}\frac{(-1)^{m}}{m!\Gamma(m+\alpha+1)}\bigg(\frac{x}{2}\bigg)^{2m+\alpha}
\end{equation}
We truncate the series at second order in distortion amplitudes,
obtaining 
\begin{widetext}
\begin{eqnarray}
b_{n,m}(a_{E}k_{1s}) & \thickapprox & \grave{b}_{n,m}\prod_{\mu=0}^{\infty}\left[1-\frac{1}{4}\left(\left(\zeta_{E}p_{1}(k_{\mu})\right)^{2}+\left(a_{E}B_{2}p_{2}(k_{\mu})\right)^{2}\right)\right]+\mathcal{O}(p_{1}^{3},p_{2}^{3}).
\end{eqnarray}
\end{widetext}

\bibliographystyle{apsrev4-1}
\bibliography{EEHG}

\end{document}